%% file: main.tex
\documentclass[12pt,reqno]{amsart}

\usepackage{main}
\graphicspath{{figures/}}
\usepackage{titletoc}

\usepackage[style=apa,backend=biber,natbib=true,uniquename=false]{biblatex}
\newgeometry{margin=1.25in}
\hypersetup{
  pdftitle={How Common Are Estimated Latent-Distribution Departures From Normality?},
  pdfauthor={JoonHo Lee},
  pdfsubject={Latent-distribution diagnostics in item response theory},
  pdfkeywords={item response theory, latent trait distribution, nonnormality, empirical histogram}
}

\title{\Large How Common Are Estimated Latent-Distribution Departures From
Normality?\\
Evidence From 504 Item-Response Data Sets}

\author{JoonHo Lee}

\date{\footnotesize August 2026. \\[0.5em]
Lee: College of Education, The University of Alabama, Tuscaloosa, AL, USA.
\texttt{jlee296@ua.edu}.}

\renewcommand{\theHequation}{\thesection.\arabic{equation}}
\renewcommand{\theHfigure}{Main.\arabic{figure}}
\renewcommand{\theHtable}{Main.\arabic{table}}

\begin{document}

\begin{abstract}
\input{frontmatter/abstract}
\end{abstract}

\maketitle

\noindent\textbf{Keywords:} item response theory; latent trait distribution;
nonnormality; empirical histogram; Item Response Warehouse

\pagestyle{plain}
\newpage

%% ==========================================================================
%% MAIN TEXT
%% ==========================================================================

\input{sections/body}

%% ==========================================================================
%% END-MATTER STATEMENTS
%% ==========================================================================

\bigskip

\noindent\textbf{Acknowledgments.}\enspace
The author thanks the Item Response Warehouse team for building and
maintaining the public data resource on which this study depends.

\smallskip

\noindent\textbf{Funding and competing interests.}\enspace
This research received no specific grant from any funding agency in the
public, commercial, or not-for-profit sectors. The author has no conflicts of
interest to disclose.

\smallskip

\noindent\textbf{Software and reproducibility.}\enspace
The analyses used R 4.6.0 \parencite{r_core_team_r_2026}, \pkg{mirt} 1.46.1
for normal, empirical-histogram, extrapolated-empirical-histogram, and
Davidian fits \parencite{chalmers_mirt_2012}, \pkg{diptest} 0.77.2
\parencite{maechler_diptest_2025}, \pkg{clubSandwich} 0.7.0
\parencite{pustejovsky_clubsandwich_2026}, \pkg{data.table} 1.18.4
\parencite{barrett_datatable_2026}, and \pkg{ggplot2} 4.0.3
\parencite{wickham_ggplot2_2016}. The analysis contract, sample-flow
membership, estimator settings, random seeds, software versions, response
hashes, and failure states are stored with the replication package. Code,
manifests, derived unit-level summaries, and the source map for reported
results are released at
\mbox{\url{https://github.com/joonho112/irw-normality-replication}}. The
study results can also be explored interactively at
\mbox{\url{https://joonho112.github.io/irw-normality-webapp/}}.

\smallskip

\noindent\textbf{Data availability.}\enspace
All item-response data analyzed here are public through the Item Response
Warehouse \parencite[IRW;][]{domingue_introduction_2025},
\url{https://datapages.github.io/irw/}. The analyzed frame contains 504
response matrices from 273 studies. Original response-level data remain
governed by the licenses of the primary IRW sources.

%% ==========================================================================
%% REFERENCES
%% ==========================================================================

\printbibliography[title={References}]

%% ==========================================================================
%% ONLINE SUPPLEMENT
%% ==========================================================================

\newpage
\appendix
\setcounter{section}{0}
\renewcommand{\thesection}{\Alph{section}}
\numberwithin{equation}{section}
\numberwithin{figure}{section}
\numberwithin{table}{section}

%% Appendix counters and hyperref anchors must remain distinct from main-text
%% figure and table counters.
\renewcommand{\theHequation}{Supp.\thesection.\arabic{equation}}
\renewcommand{\theHfigure}{Supp.\thesection.\arabic{figure}}
\renewcommand{\theHtable}{Supp.\thesection.\arabic{table}}
\crefname{section}{Appendix}{Appendices}
\Crefname{section}{Appendix}{Appendices}

\begin{center}
{\Large\bfseries Online Supplement}\\[1em]
{\large How Common Are Estimated Latent-Distribution Departures From
Normality?\\
Evidence From 504 Item-Response Data Sets}\\[1.5em]
{\normalsize JoonHo Lee}
\end{center}

\vspace{1.5em}

\noindent This supplement accompanies the main text. Section, equation,
table, and figure numbers below are prefixed by the appendix letter (A.1,
B.1, \dots). References to the main text use its unprefixed numbering.

\vspace{1.5em}

\startcontents[appendices]
\printcontents[appendices]{}{1}{\textbf{Contents}\vskip1em\hrule\vskip1em}
\vskip1em\hrule\vskip2em

\input{appendices/appendix_A}
\input{appendices/appendix_B}
\input{appendices/appendix_C}
\input{appendices/appendix_D}
\input{appendices/appendix_E}
\input{appendices/appendix_F}

\end{document}

%% file: frontmatter/abstract.tex
%% abstract.tex --- applied-reader abstract (final arXiv version).
Item response models usually assume a normal trait distribution, yet little is
known about how often fitted distributions in real studies differ substantially
from normality or which reported results are most affected. We analyzed 504
item-response data sets from 273 studies in the Item Response Warehouse,
fitting each with the normal assumption and with a flexible distribution
estimated from the responses, and we compared reliability, item estimates,
predicted test responses, and person scores between the two calibrations. In
more than half of the data sets the two fitted distributions differed by at
least 10 percentage points of cumulative probability at some point on the
trait scale, with a median maximum difference of about 11 points; about
one-third reached 15 points, and nearly one-fifth reached 20 points. The
estimated shapes included skewness, heavy tails, flat regions, and occasional
multimodality. Differences were larger in several attitudinal, affective, and
behavioral domains, although these contrasts weakened when data sets were
compared within item-model families. Reliability usually changed little when
item estimates were held fixed, whereas refitting the full model produced
larger changes in some data sets, and item estimates, predicted test
responses, and person scores did not change in parallel. Alternative flexible
methods generally identified the same data sets as most unusual but disagreed
somewhat about magnitude. Applied analyses should state the distributional
assumption, examine a flexible alternative, and report sensitivity separately
for each result used in interpretation or decision making.

%% file: sections/body.tex
%% sections/body.tex --- main text for the combined arXiv build.

\section{Introduction}\label{sec:introduction}
\input{sections/01_introduction}

\input{sections/02_method}

\input{sections/04_results}

\input{sections/05_discussion}

%% file: sections/01_introduction.tex
%% 01_introduction.tex --- no heading for the introduction itself.
%% Voice: Rabe-Hesketh & Skrondal (2006). Flat declaratives, plain
%% attribution, hedged claims, descriptive headings, plan-of-the-paper close.

Item response theory (IRT) models are usually estimated by marginal maximum
likelihood, in which the response likelihood is integrated over an assumed
distribution for the latent trait $\thetav$. The standard normal distribution
is the conventional choice \parencite{bock_marginal_1981}, and it is the
default in widely used software \parencite{chalmers_mirt_2012}. The same
distribution reappears at the scoring stage, since expected a posteriori
(EAP) scores use it as a prior, and it is inherited by scale scores, norm
tables, and cut scores derived from them. The choice of latent distribution
can therefore affect item calibration, population summaries, and person
scores.

Whether the normal choice is appropriate in a given application is an
empirical question. Normality is seldom implied by a substantive theory of
the construct. \textcite{samejima_departure_1997} argued that departures from
normality can carry substantive information rather than being a nuisance. The
populations that psychologists and educators sample are
also often selected in ways that would be expected to distort a distribution:
clinical intakes, volunteer panels, and groups routed into a program by prior
attainment. \textcite{woods_item_2009} note that nonnormality could result
from the mixing of heterogeneous populations or from selection processes
unknown to the analyst. Flexible latent-density methods have been available
for several decades, yet there is little evidence on the estimated shapes
that they return across large collections of empirical item-response data.
The present study provides such a description for 504 data sets, together
with an account of which reported quantities are sensitive to the choice.

\subsection{Observed and Latent Distributions}

For observed test scores the corresponding question has been asked
repeatedly, and for a long time. \textcite{lord_survey_1955} surveyed the
skewness and kurtosis of test-score distributions, and
\textcite{cook_replication_1959} replicated the survey.
\textcite{micceri_unicorn_1989} examined 440 large-sample achievement and
psychometric distributions, rejected formal normality for every one of them
at the .01 level, and judged 19 (4.3\%) to be reasonable approximations to
the Gaussian. \textcite{blanca_skewness_2013} classified 693 small-sample
distributions by skewness and kurtosis and found 5.5\% close to the values
expected under normality. \textcite{cain_univariate_2017} reported
significant skewness or kurtosis in almost 74\% of the distributions they
assessed and documented how rarely these quantities are reported at all.
\textcite{ho_descriptive_2015} characterized 504 raw and scale-score
distributions from state testing programs, and the same concern continues to
be raised \parencite{ferr_normal_2025}. Two features of this record seem
worth noting. Departures from normality are common whenever someone looks for
them, and the looking has been confined to sums and scale scores, which can
be inspected without fitting a model.

An observed-score distribution, however, is determined jointly by the latent
distribution, the item response functions, the scoring transformation, and
the sample, so it does not provide a direct estimate of latent shape. The
relation between the two levels is closest in \textcite{ho_descriptive_2015},
who attributed characteristic irregularities in operational score
distributions, a balancing of skewness, an increase in kurtosis, and unusual
patterns of discreteness, to the scaling and scoring procedures of IRT. Those
procedures conventionally assume a normal latent distribution. The present
analysis estimates the latent distribution with that restriction relaxed, and
so concerns the other side of the same scaling step: the distribution the
procedures assume rather than the scores they produce. Because scaling
reshapes distributions, observed shape neither implies nor rules out a
departure at the latent level, and neither analysis substitutes for the
other. Our practical conclusion is nevertheless the one that
\textcite{ho_descriptive_2015} reached from their side, that it is the
researcher's responsibility to ``fit the model to the data, not the data to
the model.''

\subsection{Estimation of Latent Distributions}

The latent distribution cannot be inspected directly, and the literature on
it is therefore largely methodological. \textcite{mislevy_estimating_1984}
formalized marginal maximum likelihood estimation with an unknown latent
distribution. In the empirical-histogram approach, masses on a quadrature
grid are re-estimated within the EM algorithm of
\textcite{bock_marginal_1981}; \textcite{woods_empirical_2007} studied this
approach for ordinal item-response models. Smooth semiparametric alternatives
have been constructed with splines and Ramsay curves
\parencite{woods_item_2006,woods_ramsaycurve_2006,woods_ramsay_2007}, with
extensions to the three-parameter logistic model
\parencite{woods_ramsaycurve_2008} and to stochastic estimation methods
\parencite{monroe_estimation_2014}, with Johnson curves
\parencite{van_den_oord_estimating_2005}, and with Davidian curves
\parencite{woods_item_2009}. Bayesian and nonparametric formulations provide
further alternatives \parencite{duncan_nonparametric_2008,
sanmartin_bayesian_2011,zhang_bayesian_2021}. A related tradition in
large-scale assessment estimates characteristics of latent populations
directly from sparse item responses, without accurate individual scores
\parencite{mislevy_estimating_1984,mislevy_population_1992}, which is a
reminder that the latent distribution is itself a reporting target and not
only a technical device.

A second strand develops tests of the latent normality assumption
\parencite{li_summed_2018,monroe_testing_2021,guastadisegni_generalized_2025}.
A third uses simulation to quantify the consequences of ignoring
nonnormality: \textcite{sass_estimating_2008} found that nonnormal latent
distributions increased the error of trait estimates while item parameters
retained comparable precision, \textcite{finch_rasch_2016} compared Rasch
estimation methods under nonnormal traits, \textcite{wang_robustness_2018}
compared full-information and limited-information estimators with skewed
latent dimensions, and \textcite{manapat_examining_2022} mapped the degrees
of nonnormality at which parameter recovery for the graded response and
two-parameter models becomes unacceptable. In an applied demonstration,
\textcite{soland_how_2024} showed that calibration and scoring decisions,
including the treatment of the latent distribution, can change the
substantive conclusions of survey-based studies.

These literatures establish methods for estimation, testing, and sensitivity
analysis. What they do not contain, as far as we are aware, is a description
of the estimated departures themselves across a large and heterogeneous
collection of real calibrations: estimator papers demonstrate their methods
on a few empirical examples, test papers establish power by simulation, and
robustness studies postulate the nonnormality whose typical size in practice
is unknown. The obstacle has been practical rather than conceptual, since
density-estimating calibration of hundreds of heterogeneous data sets
requires a harmonized corpus, and one has only recently become available.
The present study applies the estimation methods of this literature at the
scale of the observed-score surveys.

\subsection{Scope of Estimated Latent Shape}

A description of this kind requires care about what is being estimated. With
finitely many items, the response patterns impose finitely many integral
constraints and do not uniquely identify an unrestricted population mixing
distribution \parencite{sanmartin_bayesian_2011}. We therefore study the
standardized empirical-histogram diagnostic returned by a stated item model,
quadrature rule, optimizer, and density estimator. Its departure from a
matched normal distribution describes what changes when the normality
restriction is relaxed in the fitted model, which is the quantity an applied
analyst faces when choosing between the two calibrations. It does not
identify an estimator-free population density, and differences between
calibrations are accordingly interpreted as sensitivity measures rather than
as bias with respect to a known truth.

\subsection{The Present Study}

The Item Response Warehouse (IRW) provides harmonized response matrices that
vary widely in construct, sample size, test length, and response format
\parencite{domingue_introduction_2025}. We analyze 504 matrices from
273 studies. Each unit is calibrated under a normal latent distribution and
under an empirical-histogram distribution on a common standardized metric,
and the estimated shape is summarized by a grid-boundary
Kolmogorov--Smirnov (KS) distance, skewness, kurtosis, and Hartigan's dip
statistic. We denote the first quantity by $\KS$. Like classical KS, it is the
largest absolute gap between two cumulative distribution functions (CDFs).
Unlike classical KS, however, its maximum is taken only over the fixed midpoint
boundaries between adjacent EH grid cells.

We ask three questions. How large are the estimated departures from
normality across the corpus, and how often do they exceed magnitudes an
applied reader would consider sizable? How does the estimated departure vary
with the construct being measured and with features of the test? And which of
the quantities that are reported in practice, reliability, item parameters,
test response curves, and person scores, change when the restriction is
relaxed?

The plan of the paper is as follows. The Method section describes the corpus,
the two calibrations, the shape diagnostics, the relation of the design to
formal tests of latent normality, the between-calibration comparisons, and
the estimator and recovery checks. The Results section reports the magnitude
and frequency of estimated departures, their variation across constructs and
item-model families, the corresponding calibration sensitivities, and the
dependence of the diagnostic on the density estimator. Full panels for the
supporting analyses are given in the online supplemental materials (OSM;
Appendices \ref{app:notation}--\ref{app:repro}). We close with implications
for the analysis and reporting of latent distributions.

%% file: sections/02_method.tex
%% 02_method.tex --- V1.9 Method. Voice: Rabe-Hesketh & Skrondal (2006).
%% Reviewed V1.9 numbers are enforced by tools/verify_v18_numbers.R.

\section{Method}

\subsection{Corpus}

We analyzed response matrices from the versioned Item Response Warehouse (IRW)
described by \textcite{domingue_introduction_2025}, an open repository that
distributes contributed item-response data in a harmonized person-by-item
format. A unit was one instrument administered to one sample. Units from the
same contributed data set were treated as belonging to the same study, and the
study is the clustering level in all pooled analyses.

The dispatcher attempted 590 candidate units. Worker-level extraction and
screening succeeded for 572 units, and 504 units from 273 studies yielded a
valid empirical-histogram (EH) shape diagnostic under the revised analysis
contract. Eight of the 504 units did not have the complete pair of normal and
EH calibrations needed for the sensitivity comparisons, leaving 496 pairs.
Membership and failure status at each stage were recorded in a unit-level
manifest, so no unit is silently absent from a reported denominator.

Eligibility required at least 500 usable persons, three items, and two
observed response categories. For units with more than 5,000 persons we drew
one deterministic, seeded subsample of 5,000. The target of the analysis is
the shape of an estimated population distribution rather than any individual
score, and 5,000 observations are ample for the summaries used here; the cap
also keeps the repeated refitting required by the estimator comparisons
feasible. Among the analyzed units, persons ranged from 500 to 5,000 (median
1,410.5), items from 3 to 379 (median 19), and the maximum number of response
categories from 2 to 11 (median 5). \Cref{tab:corpus} summarizes the flow and
composition.

Dichotomous units were fitted with a Rasch model
\parencite{rasch_probabilistic_1960}, and polytomous units were fitted with a
generalized partial credit model \parencite[GPCM;][]{muraki_generalized_1992};
there were 120 Rasch and 384 GPCM units. The reason for constraining the
dichotomous fits is identification. Estimating free item discriminations and
a free latent density from the same short dichotomous test asks a great deal
of the data, and the combination can be weakly identified; the GPCM retains
free discriminations for the polytomous units, where the additional response
categories carry more information per item. The IRW is a collection of data
that could be shared, not a probability sample of instruments or studies, and
the frequencies reported below describe the analyzed units.

\input{floats/tab1_corpus}

\subsection{Calibrations and Estimated Distribution}

Each unit was fitted twice by marginal maximum likelihood in \pkg{mirt}
\parencite{chalmers_mirt_2012}. The first calibration used \pkg{mirt}'s
Gaussian mixing distribution. Its raw fitted coordinate was not assumed to be
uniformly $N(0,1)$ across item-model families: in particular, the Rasch
parameterization frees the latent variance for scale identification. The second
calibration estimated the masses on a 121-point
quadrature grid within the EM algorithm, the empirical-histogram method of
\textcite{bock_marginal_1981} in the ordinal-data implementation studied by
\textcite{woods_empirical_2007}. Both fits used the same response matrix,
item model, item-key decisions, and optimizer controls, so the two
calibrations differ only in the latent distribution. For every cross-calibration
shape, item, and score comparison, we transformed each fitted latent coordinate
post hoc to weighted mean zero and variance one. On that common metric, the EH
diagnostic varies in shape rather than in arbitrary location or scale.

Two conversions place the item comparisons on a common footing. For Rasch
units, we retained the common-slope constraint and linked both fits to the
same standardized scale before comparing parameters. For GPCM units, category
intercepts were converted to ordered threshold locations and then
standardized. Without these steps, a change in the latent scale or in the
intercept parameterization could be mistaken for a change in item location.

The estimated distribution is conditional on this item model, density
estimator, grid, and optimization rule. A finite item set supplies finitely
many response-pattern constraints and does not identify an unrestricted
population mixing distribution \parencite{sanmartin_bayesian_2011}. We
therefore use the standardized EH estimate as a common diagnostic of what is
obtained when the normal restriction is relaxed under the stated calibration,
and we interpret differences between calibrations as sensitivity rather than
as recovery of a true density. Grid and item-model checks are reported in the
online supplemental materials (OSM).

\subsection{Shape Diagnostics}

The primary diagnostic was the grid-boundary KS distance, denoted by $\KS$.
Classical one-sample KS maximizes the absolute CDF gap over all real $t$,
whereas $\KS$ maximizes only over the fixed EH cell boundaries defined below.
Let
$B_0=-\infty$, $B_K=\infty$, and let each interior $B_k$ be the midpoint
between adjacent standardized EH nodes. If
$W_k=\sum_{j\leq k}w_j$ and $W_0=0$, we computed
\[
  \KS=\max_{0\leq k\leq K}\lvert W_k-\Phi(B_k)\rvert.
\]
Because the estimate is standardized, the matched normal is the standard
normal, and $\KS$ can be read directly as a discrepancy in cumulative mass: a
value of .10 means that at one evaluated cell boundary the estimate places ten
percentage points more, or less, of its mass below that boundary than the
normal distribution implies. This boundary functional avoids the spurious
contribution that arises when a whole grid mass is assigned to a single node.
Because the EH distribution is model-estimated and the maximum is restricted
to the declared boundaries, $\KS$ is a descriptive distance rather than a
classical KS test statistic; classical critical values and $p$-values do not
apply (OSM Appendix~\ref{app:notation}).

Signed skewness and excess kurtosis, computed directly from the standardized
grid masses, describe the direction and form of a departure. Multimodality was
indexed by a Monte Carlo Hartigan-dip descriptor
\parencite{hartigan_dip_1985}. We drew 4,000 observations from the standardized
node masses with replacement, added Gaussian jitter with SD .05 on that
standardized scale, and computed the dip with a deterministic per-unit seed.
The result is a lightly smoothed continuous descriptor, not an exact dip of the
unsmoothed masses. Plotted densities were smoothed separately with a common
kernel for legibility.

We summarized the complete empirical distribution of KS and its exceedance
function. The shares above .05, .10, .15, and .20 provide readable landmarks
on that function.

\subsection{Relation to Tests of Latent Normality}

Dedicated tests of the latent normality assumption exist, and it is worth
stating how the present design relates to them. The summed-score likelihood
indices of \textcite{li_summed_2018}, the posterior-residual index of
\textcite{monroe_testing_2021}, and the generalized Hausman test of
\textcite{guastadisegni_generalized_2025} are constructed specifically for
this assumption. The same studies report that general-purpose
limited-information fit statistics of the $M_2$ family have little power
against misspecification of the latent distribution, so a routinely
acceptable overall fit statistic provides no assurance about latent shape.
The summed-score and posterior-residual statistics are also reported to be
sensitive to distributional violations while remaining largely insensitive to
multidimensionality, which makes them attractive within their validated
ranges. The generalized Hausman test has instead been validated specifically
within the dichotomous two-parameter setting.

We did not adopt these tests as the corpus instrument. Coverage is one
reason: the generalized Hausman test is developed for the
dichotomous two-parameter model, whereas 384 of the 504 units here are
polytomous, and its published validation covers much shorter tests than this
corpus contains, with reduced power reported under extreme item slopes and
convergence problems with few items. Implementations are uneven: the
unadjusted summed-score statistic is available in \pkg{flexMIRT}, and an
R implementation is available in the \pkg{rpf} package
\parencite{pritikin_rpf_2025}. The posterior-residual index, however, is not
available in commercial software
\parencite{guastadisegni_generalized_2025}. A corpus description also requires
a magnitude on a common scale across dichotomous and polytomous units, which
the KS distance supplies and a family of test-specific statistics does not.

The exceedance shares reported below are therefore descriptive counts, not
significance decisions. A formal rejection fraction would require a
separately calibrated unit-level reference distribution for the fitted
diagnostic, and it is not reported in this study.

\subsection{Construct and Design Comparisons}

IRW metadata placed the units in eight named construct categories or an
unclassified category. We report category medians and interquartile ranges.
For adjusted comparisons, we regressed $\operatorname{logit}(\KS)$ on
construct, log items, log persons, longitudinal status, item-model family,
and the maximum number of response categories. Because several studies
contribute more than one unit, the ordinary least squares estimates were
accompanied by CR2 standard errors clustered by study with Satterthwaite
degrees of freedom \parencite{pustejovsky_clubsandwich_2026}. Interaction and
model-family-specific fits were used to assess whether the pooled construct
pattern was supported within the Rasch and GPCM families separately.

\subsection{Between-Calibration Sensitivity}

For the 496 complete pairs we compared the two calibrations on the quantities
that applied reports commonly use. Neither calibration is known to be true,
so the differences are reported as between-calibration sensitivity, not as
bias.

We distinguished three reliability-related targets because they isolate
different sources of change. For the fixed-item density component, the
normal-calibration item parameters were held fixed and the bounded information
ratio
\[
  \rho(\thetav)=\frac{\Jinfo(\thetav)}{\Jinfo(\thetav)+1}
\]
was integrated once over the normal density and once over the EH density. This
descriptive transformation places test information on a reliability-like
$[0,1)$ scale; it is not the conditional or marginal reliability coefficient
defined by \textcite{green_technical_1984}. The comparison isolates the change
due to reweighting alone. Full-refit marginal reliability was computed after
the item parameters and the density had both been re-estimated, so it combines
reweighting with item-parameter change. Full-refit empirical reliability was
computed from the scores returned by each complete calibration. We report the
absolute difference for each target.

Item comparisons used the root mean square difference (RMSD) in standardized
Rasch locations or GPCM thresholds and, where applicable, standardized
discriminations. We also calculated the RMSD between test characteristic
curves and divided it by the number of items; this summarizes what the
parameter differences imply for predicted test scores. Person comparisons
used the mean absolute difference between standardized EAP scores, each
computed under its own calibration and its own prior density, and the
fraction of persons whose classification relative to $|z|=1$ changed between
fits. The boundary is a descriptive reference, approximately the 16th and
84th percentiles under normality, of the kind used to define screening
regions.

The item and score distributions have long right tails, which contain poorly
aligned or unstable refits; medians and quartiles are the main summaries.

\subsection{Model, Estimator, and Recovery Checks}

Eight units spanning item-model family, test length, and KS were refitted
under seven grid and optimizer configurations. The same dichotomous units
were also fitted with Rasch, 2PL, and, where stable, 3PL item response
functions. These checks measure the dependence of the diagnostic on numerical
settings and on the working item model.

We re-estimated all 504 units with the extrapolated empirical-histogram
variant implemented in \pkg{mirt} (EHW) and with Davidian curves of orders 2,
4, 6, 8, and 10. Davidian order was selected by the Hannan--Quinn criterion,
the rule prescribed by \textcite{woods_item_2009}, under candidate caps of 6
and 10. GPCM comparisons retained the same item-model parameterization. The
available Rasch Davidian specification uses a common-slope, fixed-variance
reparameterization, so its results combine density-estimator and
parameterization differences and are labeled accordingly.

A known-truth simulation, reported following the structure recommended by
\textcite{siepe_simulation_2024}, crossed four standardized latent shapes,
2PL and GPCM item families, and 10- and 20-item tests at $N=5{,}000$. One
hundred Monte Carlo jobs were planned for each of the 16 cells, and each job
fitted EH and the five Davidian candidates. We assessed the difference
between detected and true KS and the KS distance between the estimated and
generating distributions. The simulation concerns density recovery for the
implemented procedures; it does not supply a formal test for the corpus.

\input{sections/03_transparency}

%% file: floats/tab1_corpus.tex
\begin{table}[!htbp]
\caption{Analysis Flow and Corpus}
\label{tab:corpus}
\begin{apatable}
\centering
\begin{tabular}{lrrr}
\toprule
Stage or model family & Units & Studies & Persons/items/categories \\
\midrule
Candidate units attempted & 590 & --- & --- \\
Worker-level success & 572 & --- & --- \\
EH diagnostic analysis frame & 504 & 273 & --- \\
Complete calibration pairs & 496 & --- & --- \\
\addlinespace
Rasch & 120 & 86 & 1,914 / 45 / 2 \\
GPCM & 384 & 196 & 1,375 / 12 / 5 \\
\bottomrule
\end{tabular}
\end{apatable}
\tablenote{The last column gives medians for persons used, items used, and
maximum response categories. Across all 504 units, the corresponding medians
were 1,410.5, 19, and 5; ranges were 500--5,000 persons, 3--379 items, and
2--11 categories. Study counts by model family overlap because some studies
contributed both families. EH = empirical histogram; GPCM = generalized partial
credit model.}
\end{table}

%% file: sections/03_transparency.tex
\subsection{Transparency and Openness}

The item-response data are public through the IRW. The analysis contract,
sample-flow membership, failure states, estimator settings, random seeds,
software versions, and source hashes are stored with the replication package.
All article tables and figures are generated from the public unit-level
analysis dataset and documented release outputs. Code, manifests,
derived unit-level summaries, and the source map for reported results are
released at \url{https://github.com/joonho112/irw-normality-replication}.
The study was not preregistered. Analytic decisions are
recorded in a dated log. There was no random assignment or intervention, and
the analyses do not estimate causal effects
\parencite{appelbaum_journal_2018}.

The analyses used R 4.6.0 \parencite{r_core_team_r_2026}, \pkg{mirt} 1.46.1
for normal, EH, EHW, and Davidian fits \parencite{chalmers_mirt_2012},
\pkg{diptest} 0.77.2 \parencite{maechler_diptest_2025}, \pkg{clubSandwich}
0.7.0 \parencite{pustejovsky_clubsandwich_2026}, \pkg{data.table} 1.18.4
\parencite{barrett_datatable_2026}, and \pkg{ggplot2} 4.0.3
\parencite{wickham_ggplot2_2016}. \pkg{sirt} was used only for an archived
\code{smooth3} comparison \parencite{robitzsch_sirt_2025}.

%% file: sections/04_results.tex
%% 04_results.tex --- V1.9 Results. Corrected dip summaries; other targets retained.

\section{Results}

\subsection{Estimated Latent Distributions}

Of the 590 attempted response matrices, 504 from 273 studies yielded an EH
shape diagnostic. They comprised 120 Rasch and 384 GPCM units, and the median
analyzed unit contained 1,410.5 persons and 19 items (\cref{tab:corpus}).

\begin{figure}[!htbp]
\caption{Selected Estimated Latent Distributions}
\label{fig:menagerie}
\begin{apatable}
\centering
\includegraphics[width=\textwidth]{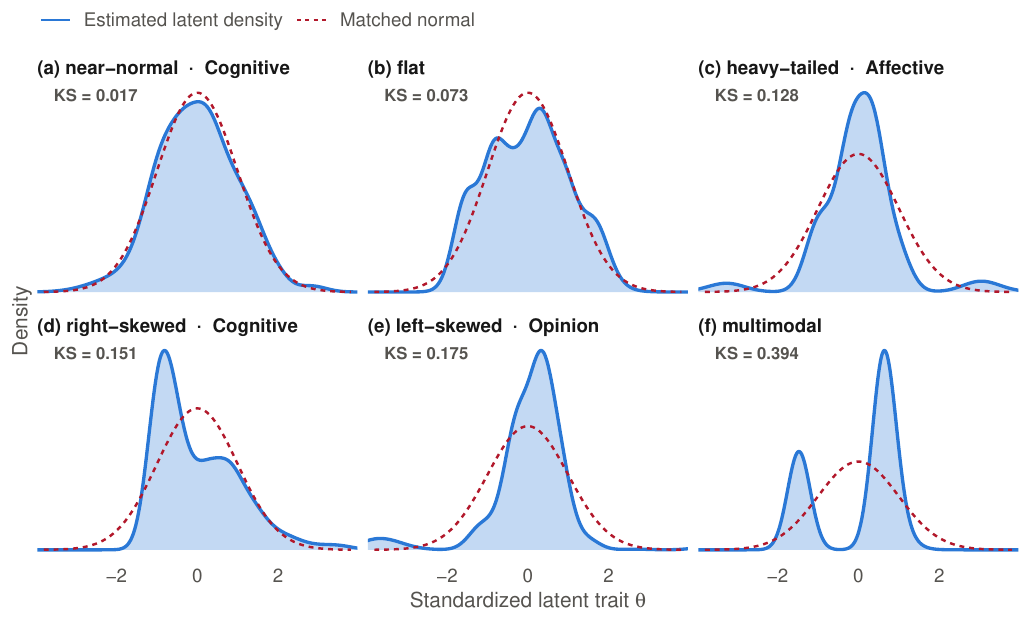}
\end{apatable}
\figurenote{The solid curves are smoothed standardized EH estimates, and the
dashed curves are standard normal densities. The six units were selected to
show the principal forms observed in the revised analysis frame: (a) a
near-normal cognitive unit, $\KS=.017$; (b) a flat unit, $.073$; (c) a
heavy-tailed affective unit, $.128$; (d) a right-skewed cognitive unit,
$.151$; (e) a left-skewed opinion unit, $.175$; (f) a multimodal unit,
$.394$. Panel labels give the grid-boundary KS distance calculated from the
unsmoothed EH masses. Panels have separate vertical scales and compare shape
rather than density height. EH = empirical histogram; KS = grid-boundary
Kolmogorov--Smirnov distance.}
\end{figure}

\Cref{fig:menagerie} shows six standardized EH estimates selected to
illustrate the main shapes in the corpus, ordered by their KS distance. The
first, a cognitive unit with $\KS=.017$, is nearly indistinguishable from the
normal curve. The second, at $.073$, is flatter near the center than the
normal density. The remaining panels depart more clearly: an affective unit
with heavy tails at $.128$, right- and left-skewed units at $.151$ and
$.175$, and a unit at $.394$ in which the estimate has two separated modes
where the normal calibration places a single central peak. A unit at the
corpus median resembles the second and third panels rather than the first.
These curves are estimated under their working item models. In particular,
separated modes may reflect population mixtures, selection, local dependence,
or model misspecification; the plot does not distinguish among these sources.

The median KS distance was .109, with an interquartile range of .070 to .166 and a
90th percentile of .257 (\cref{tab:shape}). Signed skewness had a median of
.178, so neither direction of asymmetry dominated the corpus, whereas its
absolute value had a median of .799. Median excess kurtosis was 3.584. The
dip descriptor had a median of .018 and a 90th percentile of .055, so
pronounced multimodality was a minority feature.

\input{floats/tab2_prevalence}

\begin{figure}[!htbp]
\caption{Magnitude and Frequency of Estimated Departures}
\label{fig:prevalence}
\begin{apatable}
\centering
\includegraphics[width=\textwidth]{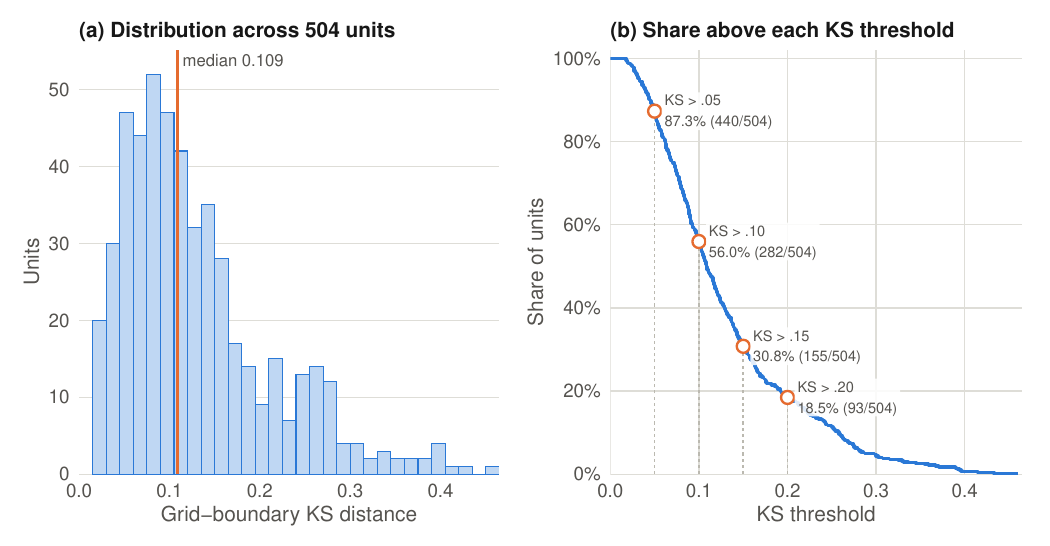}
\end{apatable}
\figurenote{The left panel shows the grid-boundary KS distance for all 504 units; the
vertical line marks the median. The right panel shows the empirical
proportion of units above each KS value. Labels give counts and percentages
at four descriptive landmarks. No cutoff is a significance threshold.}
\end{figure}

The empirical exceedance curve in \cref{fig:prevalence} gives the frequency
of departures without selecting a single cutoff. KS exceeded .05 in 440 units
(87.3\%), .10 in 282 units (56.0\%), .15 in 155 units (30.8\%), and .20 in 93
units (18.5\%). More than half of the analyzed units therefore had a maximum
CDF difference greater than ten percentage points, and nearly one unit in
five had a difference greater than twenty. These are descriptive corpus
frequencies, not test rejections.

\subsection{Construct and Item-Model Variation}

\begin{figure}[!htbp]
\caption{KS Distance by Construct and Item-Model Family}
\label{fig:construct}
\begin{apatable}
\centering
\includegraphics[width=\textwidth]{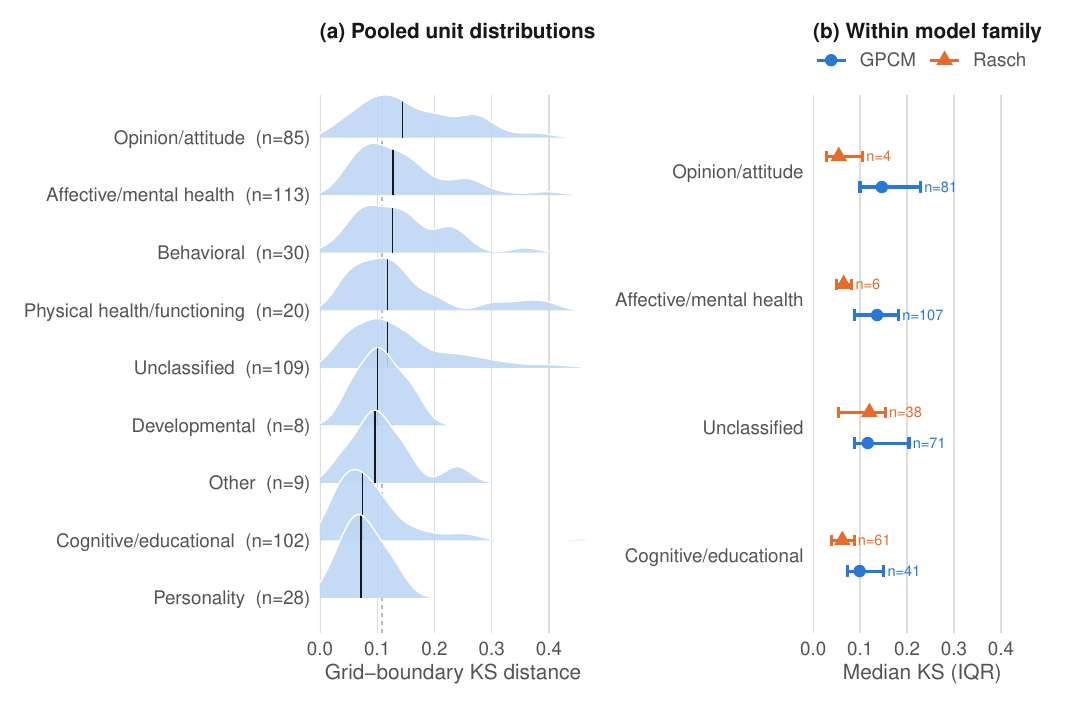}
\end{apatable}
\figurenote{The left panel shows unit-level distributions and pooled medians.
The right panel shows model-family-specific medians and interquartile ranges;
labels give the number of units. Sparse cells are shown to document the
available support and should not be treated as stable summaries.}
\end{figure}

Pooled median KS ranged from .071 for personality and .074 for
cognitive/educational units to .128 for affective/mental health and .144 for
opinion/attitude units, with behavioral units at .127. The distributions
overlapped substantially within every construct (\cref{fig:construct}).
Construct and response format are also entangled in this corpus: 81 of 85
opinion/attitude units and 107 of 113 affective/mental health units used the
GPCM, whereas 61 of 102 cognitive/educational units used the Rasch model.
Any comparison of constructs is therefore partly a comparison of item-model
families.

\input{floats/tab3_metareg}

The additive model in \cref{tab:metareg} adjusts the pooled contrasts for
design features. Relative to cognitive/educational units, the
affective/mental health and opinion/attitude coefficients on the logit KS
scale were .398 ($p=.007$) and .444 ($p=.005$). The coefficient for log items
was $-.170$ ($p=.001$), so estimated departures were smaller in longer
tests, and the coefficient for log persons was small ($-.046$, $p=.459$).
The GPCM indicator was .036 ($p=.817$) after adjustment. In the GPCM-only
model, however, the two construct coefficients fell to .221 ($p=.193$) and
.252 ($p=.150$). The pooled pattern therefore describes this corpus; it does
not establish a construct ordering that is invariant to the item-model
family (OSM Appendix~\ref{app:robust}).

\subsection{Between-Calibration Sensitivity}

The three reliability targets gave different results among the 496 complete
calibration pairs (\cref{tab:sensitivity}). The median absolute difference
was .002 when the item parameters were held fixed and only the integration
density was changed, with 1.0\% of units differing by more than .05. It was
.025 for full-refit marginal reliability, with 27.2\% above .05, and .005 for
full-refit empirical reliability, with 6.0\% above .05. Density reweighting
alone therefore changed reliability little in most units; the larger
full-refit differences also include the change in item parameters, and the
three targets answer different questions about the same calibration choice.

\input{floats/tab4_bias}

Item and score estimates moved more than the fixed-item reliability
component. The median standardized threshold or location RMSD was .274, and
the median discrimination RMSD was .161. The median RMSD between test
characteristic curves, which aggregate the item parameters into predicted
test scores, was .077 per item. Standardized person scores had a median absolute
difference of .093, and a median 4.8\% of persons changed classification
relative to the $|z|=1$ reference. The threshold/location distribution was
particularly long-tailed, with 75th and 90th percentiles of .751 and 10.371;
this tail contains poorly aligned or unstable refits, and the medians are the
appropriate summaries.

\begin{figure}[!htbp]
\caption{Between-Calibration Sensitivity and KS Distance}
\label{fig:sensitivity}
\begin{apatable}
\centering
\includegraphics[width=\textwidth]{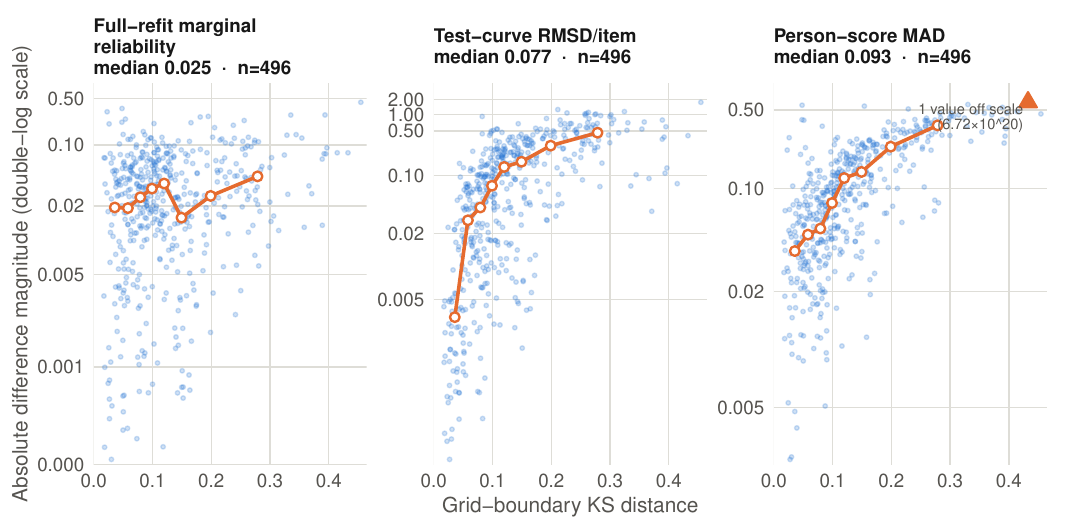}
\end{apatable}
\figurenote{Panels show full-refit marginal reliability, test-curve RMSD per
item, and standardized person-score mean absolute difference for 496 complete
normal--EH calibration pairs. Blue points are units; orange points and lines
are medians within KS octiles. The vertical display scale is transformed to
show the right tails. One person-score value of $6.72\times10^{20}$ is marked
at the display limit of .60 with an upward triangle; all raw values are
retained in the source table. No other values are capped.}
\end{figure}

\Cref{fig:sensitivity} plots three of these outcomes against KS. The blue
points show substantial dispersion at similar KS values, and the orange
summaries show the median outcome within KS octiles. KS describes a
difference between two estimated distribution functions; it does not by
itself determine how much a particular reported quantity will change in a
particular unit.

\subsection{Model and Density-Estimator Dependence}

Across the eight-unit grid study, 54 of 56 configuration fits converged, and
the within-unit range of KS across settings varied from .002 to .055.
Changing a dichotomous unit from Rasch to 2PL changed KS by a median absolute
value of about .015 among complete pairs, with larger and less even changes
in some 3PL fits. The full results (OSM Appendix~\ref{app:robust}) show that the EH
diagnostic can be numerically stable while remaining conditional on the
working item model.

\input{floats/tab5_robustness}

EH and EHW estimates were strongly associated in the 429 complete pairs
(Spearman $\rho=.876$), with a median absolute KS difference of .009
(\cref{tab:estimator}). The GPCM comparison between EH and the cap-10
Davidian procedure gave $\rho=.790$ and a median absolute difference of
.042, and the two Davidian candidate caps were closely related to each other
($\rho=.935$). Agreement at the descriptive .10 landmark was 87.9\% for EH
and EHW and 75.8\% for EH and the cap-10 Davidian procedure. The broad
ordering of units was therefore reproducible across related estimators,
whereas exact magnitudes depended in part on the density family and its
tuning rule (OSM Appendix~\ref{app:estimator}).

In the known-truth simulation (OSM Appendix~\ref{app:recovery}), EH
detected-minus-true KS was .002 for 2PL data and .004 for GPCM data in the
20-item bimodal cells. The corresponding values for the
Hannan--Quinn-selected Davidian procedure were $-.042$ and $-.049$ when the
candidates extended to order 10, and $-.099$ and $-.093$ under the cap of 6,
so raising the ceiling reduced the shortfall. Under normal generation the
selected Davidian values were close to zero, whereas EH values were .012 and
.009. These results concern the fitted procedures and generating shapes used
here; Davidian curves can represent multimodal distributions when an
adequate candidate order is available \parencite{woods_item_2009}.

%% file: floats/tab2_prevalence.tex
\begin{table}[!htbp]
\caption{Magnitude and Frequency of Estimated Distributional Departures}
\label{tab:shape}
\begin{apatable}
\centering
\setlength{\tabcolsep}{7pt}
\begin{tabular}{lrrrrr}
\toprule
Diagnostic & \textit{P}10 & \textit{P}25 & Mdn & \textit{P}75 & \textit{P}90 \\
\midrule
Grid-boundary KS distance & .045 & .070 & .109 & .166 & .257 \\
Absolute skewness & .195 & .366 & .799 & 1.540 & 2.436 \\
Excess kurtosis & $-.122$ & .986 & 3.584 & 8.444 & 14.976 \\
Hartigan dip & .004 & .009 & .018 & .034 & .055 \\
\bottomrule
\end{tabular}

\vspace{6pt}
\begin{tabular}{lrr}
\toprule
Descriptive KS landmark & Units above landmark & Share \\
\midrule
.05 & 440 / 504 & 87.3\% \\
.10 & 282 / 504 & 56.0\% \\
.15 & 155 / 504 & 30.8\% \\
.20 & 93 / 504 & 18.5\% \\
\bottomrule
\end{tabular}
\end{apatable}
\tablenote{Each unit contributes one standardized EH estimate. KS is the
grid-boundary Kolmogorov--Smirnov distance: the maximum cumulative-mass difference at midpoint cell boundaries relative to the matched normal. Because the EH distribution is model-estimated and the maximum is restricted to those boundaries, classical KS critical values and $p$-values do not apply.
Landmark rows are points on the empirical exceedance function;
they are not formal test results. EH = empirical histogram; Mdn = median.}
\end{table}

%% file: floats/tab3_metareg.tex
\begin{table}[!htbp]
\caption{Additive Model for Logit KS Distance}
\label{tab:metareg}
\begin{apatable}
\centering
\setlength{\tabcolsep}{6pt}
\begin{tabular}{lrrrr}
\toprule
Term & $\hat\beta$ & \textit{SE} & \textit{df} & $p$ \\
\midrule
Affective/mental health & .398 & .143 & 59.1 & .007 \\
Opinion/attitude & .444 & .148 & 37.3 & .005 \\
Other/limited overlap & .213 & .129 & 40.9 & .106 \\
Unclassified & .412 & .157 & 53.8 & .011 \\
log(items) & $-.170$ & .051 & 52.9 & .001 \\
log(persons) & $-.046$ & .061 & 58.2 & .459 \\
Longitudinal & $-.002$ & .160 & 16.8 & .988 \\
GPCM indicator & .036 & .156 & 50.7 & .817 \\
Maximum categories & .041 & .036 & 20.2 & .260 \\
\bottomrule
\end{tabular}
\end{apatable}
\tablenote{$N=504$ units in 273 studies. Cognitive/educational and Rasch are
the reference levels. Ordinary least squares coefficients have CR2 standard
errors clustered by study and Satterthwaite degrees of freedom. Common-support
coding retains affective/mental health, opinion/attitude, unclassified, and a
pooled other/limited-overlap category. Model-family-specific estimates are
reported in the OSM. GPCM = generalized partial credit model.}
\end{table}

%% file: floats/tab4_bias.tex
\begin{table}[!htbp]
\caption{Absolute Between-Calibration Sensitivity in 496 Complete Pairs}
\label{tab:sensitivity}
\begin{apatable}
\centering
\setlength{\tabcolsep}{5pt}
\begin{tabular}{lrrrrr}
\toprule
Target & \textit{P}10 & \textit{P}25 & Mdn & \textit{P}75 & \textit{P}90 \\
\midrule
Fixed-item density reliability & .000 & .001 & .002 & .008 & .019 \\
Full-refit marginal reliability & .003 & .011 & .025 & .054 & .096 \\
Full-refit empirical reliability & .000 & .001 & .005 & .014 & .033 \\
Threshold/location RMSD & .041 & .105 & .274 & .751 & 10.371 \\
Discrimination RMSD & .015 & .043 & .161 & .607 & 1.653 \\
Test-curve RMSD per item & .003 & .015 & .077 & .242 & .497 \\
Standardized person-score MAD & .023 & .043 & .093 & .199 & .323 \\
$|z|>1$ reclassification & .000 & .017 & .048 & .113 & .218 \\
\bottomrule
\end{tabular}
\end{apatable}
\tablenote{Values are absolute differences between normal and EH calibrations
on a common standardized metric. They measure sensitivity to the calibration
choice. The threshold and score targets have long right tails; medians and
quartiles are the primary summaries. EH = empirical histogram; MAD = mean
absolute difference; Mdn = median; RMSD = root mean square difference.}
\end{table}

%% file: floats/tab5_robustness.tex
\begin{table}[!htbp]
\caption{Density-Estimator Dependence}
\label{tab:estimator}
\begin{apatable}
\centering
\small
\begin{tabular}{@{}lrrrr@{}}
\toprule
Comparison & $n$ & Spearman $\rho$ & Mdn $|\Delta\mathrm{KS}|$ & Agreement \\
\midrule
EH vs. EHW & 429 & .876 & .009 & .879 \\
\makecell[l]{GPCM: EH vs. Davidian HQ,\\cap 6} & 384 & .719 & .045 & .706 \\
\makecell[l]{GPCM: EH vs. Davidian HQ,\\cap 10} & 384 & .790 & .042 & .758 \\
\makecell[l]{GPCM: Davidian cap 6\\vs. cap 10} & 384 & .935 & .000 & .938 \\
\makecell[l]{Rasch reparameterization:\\EH vs. cap 6} & 120 & .572 & .029 & .750 \\
\makecell[l]{Rasch reparameterization:\\EH vs. cap 10} & 120 & .620 & .019 & .808 \\
\bottomrule
\end{tabular}
\end{apatable}
\tablenote{Agreement is classification agreement at the descriptive landmark
KS $>.10$. GPCM comparisons retain the item-model parameterization. Davidian
estimation for Rasch units used a common-slope, fixed-variance
reparameterization, so those rows also include parameterization sensitivity.
EH = empirical histogram; EHW = extrapolated empirical histogram; HQ =
Hannan--Quinn order selection; Mdn = median.}
\end{table}

%% file: sections/05_discussion.tex
%% 05_discussion.tex --- V1.9 Discussion. Voice: Rabe-Hesketh & Skrondal
%% (2006): recapitulation, interpretation, scope, caveats, practical advice.

\section{Discussion}

We have calibrated 504 item-response matrices under a normal latent
distribution and under an empirical-histogram alternative, and we have
described the estimated departures and their consequences for reported
quantities. Departures were common and often sizable. The median
grid-boundary KS distance was .109, the descriptive exceedance frequencies
were 56.0\% at .10, 30.8\% at .15, and 18.5\% at .20, and the estimated
shapes included skewness, heavy tails, flat regions, and separated modes.
Releasing the restriction left the fixed-item density component of
reliability almost unchanged in most units, changed full-refit marginal
reliability more, and moved item locations, test curves, and person scores by
amounts that varied widely across units.

\subsection{Magnitude and Frequency of Estimated Departures}

The full distribution of KS values is more informative than any single
threshold. A value of .10 means that, at one evaluated midpoint cell boundary,
the empirical-histogram estimate assigns ten percentage points more or less
cumulative mass than the matched normal distribution.
More than half of the units exceeded this value, and the corpus spans a wide
range, from a 10th percentile of .045 to a 90th percentile of .257, more
than a fivefold difference. The departures also varied in form: the median absolute skewness was
.799, the median excess kurtosis was 3.584, and the upper decile of the dip
descriptor began at .055. No single kind of departure dominated, and signed
skewness was nearly centered ($\mathrm{Mdn}=.178$), so the corpus is not
simply a collection of scales with floor or ceiling effects.

These frequencies describe the analyzed IRW corpus. They show how often the
specified diagnostic reached a given magnitude among the available units,
under the stated working models. The landmarks are descriptive and do not
define rejection rates for a formal test, and they do not estimate the
frequency of latent nonnormality in a target population of psychological and
educational instruments. Within those limits, the description seems useful:
robustness studies that must postulate a degree of nonnormality
\parencite{sass_estimating_2008,wang_robustness_2018,manapat_examining_2022}
have so far had little empirical guidance about what magnitudes are typical
of real calibrations, and the exceedance function in \cref{fig:prevalence}
provides a first anchor of that kind. The relative frequency of forms, with
asymmetry and heavy tails common and pronounced multimodality a minority,
may similarly help in choosing generating distributions for simulation
designs.

The results extend a long observed-score record. Surveys from
\textcite{lord_survey_1955} and \textcite{cook_replication_1959} through
\textcite{micceri_unicorn_1989}, \textcite{blanca_skewness_2013},
\textcite{cain_univariate_2017}, and \textcite{ho_descriptive_2015} found
departures from normality to be routine in raw and scale scores. Our
analysis concerns the latent side of the scaling step that produces such
scores: the estimated mixing distribution under an item model, on a metric
where the normal assumption is actually imposed. The two records are not
interchangeable, since scaling reshapes distributions, and an observed-score
histogram cannot determine the fitted latent distribution. That estimated
departures appear as frequently at the latent level as irregularities appear
in operational scores is therefore an empirical finding rather than a
restatement, and it is consistent with the position of
\textcite{ho_descriptive_2015} that distributional irregularity is a feature
of the data to be modeled rather than a defect to be transformed away.

\subsection{Interpretation of the Estimated Shapes}

The estimated shapes admit substantive readings, provided the
model-conditional character of the diagnostic is kept in view. Skewness and
heavy tails, the most common forms here, are what selective sampling would be
expected to produce: clinical intakes truncate the mild end of a severity
continuum, volunteer panels over-represent particular ranges of attitude, and
admission rules truncate ability distributions.
\textcite{samejima_departure_1997} argued that such departures are
substantively informative, and the present corpus suggests that occasions to
take that argument seriously arise often.

Multimodality has a sharper reading. \textcite{woods_item_2009} note that
latent nonnormality could result from the mixing of heterogeneous
populations or from selection processes unknown to the researcher, and a
bimodal estimate such as panel (f) of \cref{fig:menagerie} is what a mixture
of separated subpopulations would look like after standardization. Where
subpopulations with different response processes are suspected, mixture item
response models are available
\parencite{bolt_mixture_2001}, and questions of measurement invariance
across the latent groups arise before the scores are used. An estimated
multimodal density does not by itself establish latent subgroups, since
local dependence and item-model misspecification can produce similar
artifacts; it provides a reason to examine these possibilities, at the cost
of one additional calibration.

The construct comparisons bear on the same point, with appropriate caution.
Cognitive and educational tests were the most nearly normal group in the
pooled summaries, and attitudinal and affective scales the least. The
adjusted contrasts were clear in the pooled model but smaller and
nonsignificant within the GPCM family, and construct and response format
overlap heavily in this corpus. The available evidence therefore does not
establish a construct hierarchy that is stable across item-model families.
It does suggest that the latent-normality convention is least secure in the
survey-based literatures where scoring decisions have been shown to carry
substantive consequences \parencite{soland_how_2024}.

\subsection{Between-Calibration Sensitivity}

The calibration differences depended on the target, and the pattern has a
simple structure. With item parameters held fixed, changing the integration
density produced a median absolute reliability difference of .002, and only
1.0\% of units moved by more than .05. One reason for this stability is that
the fixed-item comparison integrates the same information ratio
$\Jinfo(\thetav)/\{\Jinfo(\thetav)+1\}$ under two densities: the ratio is
bounded and varies slowly over the region where either density has
appreciable mass, so reweighting moves mass between regions in which the
integrand differs little and the changes largely cancel. Full recalibration
gave a median absolute difference of .025 for marginal reliability, with
27.2\% of units above .05, because the item parameters and the information
function itself then change as well. The corresponding median for empirical
reliability was .005. Statements about the sensitivity of reliability should
therefore identify the target: a report that holds item parameters fixed and
one that refits the model are answering different questions, and only the
former is guaranteed to be small in this corpus.

An implication follows that seems worth stating plainly. Reliability is the
sensitivity summary that applied reports compute most routinely, and the
fixed-item component of it is the quantity least affected by the latent
distribution. A satisfactory reliability coefficient consequently provides
little assurance that the distributional convention is innocuous for other
quantities, and the near-flat left panel of \cref{fig:sensitivity} shows the
same point empirically.

Item locations, test curves, and person scores moved more. The median
standardized threshold or location RMSD was .274, a movement of about a
quarter of a trait standard deviation for the typical unit, although the
long right tail of this summary reflects unstable refits rather than typical
behavior. The median test-curve RMSD of .077 per item indicates that part of
the item-parameter movement offsets in aggregate. Person scores shifted by a
median absolute .093 standard deviations, with a median 4.8\% of persons
crossing the illustrative $|z|=1$ boundary. EAP scores are exposed through
two routes, since the latent distribution enters the likelihood during
calibration and then enters each score directly as the prior, so the two
calibrations shrink scores toward differently shaped distributions; the
resulting differences tend to be largest in the tails, where screening and
selection decisions operate. The simulation literature anticipated
this asymmetry: \textcite{sass_estimating_2008} found that nonnormal latent
distributions increased trait-estimation error while item parameters
retained comparable precision, and concluded that the additional error
arises in trait estimation. Our empirical comparisons agree and add that
reliability, item parameters, test curves, and person scores need not move
together in real data, so a single summary of calibration agreement cannot
represent all of them.

Whether differences of this size matter depends on the use. Movements of
tenths of a standard deviation may matter little for coarse group summaries,
and rank-based uses are insensitive to monotone metric changes. Uses that
attach consequences to absolute locations (cut scores, norm tables, growth
percentiles, screening boundaries) are the ones for which
between-calibration sensitivity of this size warrants a check.

\subsection{Density-Estimator Dependence}

The empirical histogram and its extrapolated variant gave similar KS
orderings ($\rho=.876$) with a median absolute difference of .009, so the
diagnostic is not an artifact of one implementation. Comparisons with
Hannan--Quinn-selected Davidian curves were less concordant ($\rho=.719$ and
.790 under caps of 6 and 10; median absolute differences .045 and .042), and
the two caps agreed closely with each other. The broad ordering of units is
therefore reproducible across related estimators, while exact magnitudes,
and the classification of borderline units against any fixed landmark,
depend on the density family and its selection rule. The family, order cap,
and selection criterion should consequently accompany any reported shape
diagnostic, in the same way that an estimator accompanies a point estimate.

The known-truth simulation clarifies the direction of these disagreements
for the procedures used here. On the selected bimodal shape, the EH estimate
recovered the generating KS nearly without error, whereas the
Hannan--Quinn-selected Davidian procedure understated it, by less when the
candidate ceiling was raised from 6 to 10; under normal generation the
selected Davidian values were nearly exact, while EH showed a small positive
excess (.012 and .009). Neither estimator dominated in every cell, and the
results are bounded by the four shapes, two item families, and selection
rule examined (OSM Appendix~\ref{app:recovery}). \textcite{woods_item_2009} showed
that Davidian curves represent complex shapes well when an adequate order is
selected, and the ceiling effect observed here is consistent with their
prescription to select among a wide range of candidate orders.

\subsection{Limitations}

The IRW is a convenience corpus of data that could be shared. The
descriptive frequencies do not generalize by design to any wider population
of instruments, the construct classifications are coarse, and although the
pooled models account for the clustering of units within studies, they
cannot account for selection into the archive.

The primary calibrations use unidimensional Rasch and GPCM working models.
Multidimensionality, local dependence, differential item functioning, and
unmodeled lower asymptotes can each leave traces in a flexible density
estimate, and the diagnostic does not separate these sources from population
shape. The item-model checks show that the diagnostic is numerically stable
under grid and optimizer variation and moves modestly between Rasch and 2PL
fits, but a working model is still a working model. No lower asymptote was
fitted here; Ramsay-curve methods that estimate a nonnormal density jointly
with guessing parameters exist \parencite{woods_ramsaycurve_2008} and would
support a sharper analysis of multiple-choice cognitive tests. Persons were
capped at 5,000 per unit, and tail detail is limited by the 121-point grid.
These design choices form part of the definition of the diagnostic.

The calibration comparisons contain no known truth, so they support
sensitivity statements about specified fitted quantities and nothing
stronger. Decision-specific claims would require an external criterion, a
consequential threshold, or a defensible data-generating mechanism for the
application at hand. For the same reason the exceedance shares are
descriptive: converting them into rejection rates would require a
per-unit reference distribution for the fitted diagnostic, calibrated under
the same estimator and design, which we regard as a natural next step rather
than a settled matter.

\subsection{Implications for Reporting Practice}

Four suggestions follow for applied work. (a) The latent-distribution
assumption should be stated, and a flexible alternative should be examined
whenever calibration or scoring may depend on it; a second calibration with
\code{dentype = "empiricalhist"} in \pkg{mirt}, or with one of the
semiparametric families, standardized to the same metric, is sufficient for
inspection. (b) The estimated shape should be reported with numerical
summaries, KS distance, skewness, excess kurtosis, and the dip statistic,
alongside the reliability coefficient, which cannot substitute for them.
(c) Sensitivity analyses should match the intended use: fixed-item
reweighting isolates the density component of a summary such as marginal
reliability, a full refit describes operational recalibration, and item,
test-curve, and person-score comparisons address different quantities and
should be reported separately, with particular attention to tail-dependent
decisions. (d) The density family, quadrature or smoothing controls, and any
order-selection rule should be named, since the magnitude of the diagnostic
depends on them; large or irregular estimated departures warrant checks of
the item model and of the composition of the sample before a substantive
reading is given.

For methodological work, two obstacles seem worth removing. The dedicated
tests of latent normality \parencite{li_summed_2018,monroe_testing_2021,
guastadisegni_generalized_2025} would see wider use, in studies of this kind
and in applications, if public implementations covering polytomous models
were available. And descriptive diagnostics of the kind reported here would
support formal per-unit inference if reference distributions calibrated
under the same estimator were developed for them. Open corpora such as the
IRW make it possible to ask how often a methodological problem actually
arises in practice, which is a question that simulation alone cannot
answer; we expect that the present description will be revised as the
warehouse and the available estimators grow.

%% file: appendices/appendix_A.tex
\section{Estimand, Scale, and Diagnostic}
\label{app:notation}

\subsection{Estimator-Defined Target}

Let $Y$ denote the responses to a finite set of items. The analysis target is
the standardized mixing distribution returned by a stated item model, density
family, quadrature rule, optimizer, and convergence rule. A finite item set
supplies only finitely many response-pattern constraints, so it does not
identify an unrestricted population mixing distribution
\parencite{sanmartin_bayesian_2011}. The empirical-histogram (EH) estimate is
therefore interpreted as a model-conditional description of latent shape.

Each unit was fitted by marginal maximum likelihood in \pkg{mirt}
\parencite{chalmers_mirt_2012}. The normal calibration used \pkg{mirt}'s
Gaussian mixing distribution; its raw scale was not uniformly fixed to
$N(0,1)$ because the Rasch parameterization frees the latent variance. The EH
calibration estimated masses on a 121-point grid
within the EM algorithm \parencite{bock_marginal_1981,woods_empirical_2007}.
Both calibrations used the same response matrix, item-model family, keying
decisions, and optimizer controls. Both fitted latent coordinates were
standardized post hoc to weighted mean zero and variance one before shape,
item, or score comparisons.

Dichotomous units were fitted with a Rasch model and polytomous units with a
generalized partial credit model (GPCM). Rasch quantities were placed on a
common mean-zero, unit-standard-deviation metric before the calibrations were
compared. GPCM category intercepts were converted to ordered threshold
locations on the same standardized metric. The item summaries in the article
therefore refer to locations or thresholds, not raw category intercepts.

\subsection{Grid-Boundary KS Distance}

Let $t_1<\cdots<t_K$ denote the standardized quadrature nodes and let $w_k$
denote the EH masses. Define $B_0=-\infty$, $B_K=\infty$, and
$B_k=(t_k+t_{k+1})/2$ for $k=1,\ldots,K-1$. With
$W_k=\sum_{j\leq k}w_j$ and $W_0=0$, we computed
\[
  \KS=\max_{0\leq k\leq K}|W_k-\Phi(B_k)|,
\]
where $\Phi$ is the standard normal distribution function. This calculation
avoids assigning all of a grid mass to its right node. Like classical KS, the
quantity is a maximum absolute CDF gap; unlike classical KS, the maximum ranges
only over the fixed set $\{B_0,\ldots,B_K\}$ rather than all real $t$. Because
the EH distribution is model-estimated, classical KS critical values and
$p$-values do not apply. The same implementation was used for the corpus,
sensitivity analyses, and known-truth recovery.

We computed signed skewness and excess kurtosis directly from the standardized
grid masses. For the Hartigan-dip descriptor
\parencite{hartigan_dip_1985}, we drew 4,000 observations from those masses,
added Gaussian jitter with SD .05 on the standardized scale, and used a
deterministic per-unit seed. Thus the dip value is a smoothed Monte Carlo
descriptor; density curves were smoothed separately for display.

\begin{table}[H]
\caption{Notation for the model-conditional shape and sensitivity analyses.}
\label{tab:osm-notation}
\centering
\small
\begin{tabularx}{\textwidth}{@{}l>{\raggedright\arraybackslash}X@{}}
\toprule
Symbol & Meaning \\
\midrule
$\thetav$ & Standardized latent coordinate under a specified calibration \\
$\geh$ & EH mixing-distribution estimate conditional on the item model and
estimation controls \\
$\hat G$ & Distribution function of the standardized EH estimate \\
$\Phi$ & Standard normal distribution function \\
$\KS$ & Grid-boundary KS distance: maximum absolute CDF gap from $\Phi$ over fixed midpoint cell boundaries \\
$\dipstat$ & Standardized resample--jitter Hartigan-dip descriptor \\
$\Jinfo(\thetav)$ & Test information under a specified calibration \\
$\rho(\thetav)$ & Conditional information-ratio reliability
$\Jinfo/(\Jinfo+1)$ \\
\bottomrule
\end{tabularx}
\end{table}

%% file: appendices/appendix_B.tex
\section{Corpus Construction and Construct Composition}
\label{app:corpus}

\subsection{Analysis Flow}

The source was the versioned Item Response Warehouse (IRW) snapshot described
by \textcite{domingue_introduction_2025}. A unit was one response matrix. The
dispatcher attempted 590 candidate units. Worker-level extraction and
screening succeeded for 572 units, and 504 units yielded a valid EH shape
diagnostic under the revised analysis contract. These units came from 273
studies. A complete normal--EH calibration pair was available for 496 units;
the remaining eight contributed to shape summaries only.

Eligibility required at least 500 usable persons, three items, and two
observed response categories. Units with more than 5,000 persons were analyzed
with a deterministic seeded subsample of 5,000. Among the 504 analysis units,
the number of persons ranged from 500 to 5,000 (median 1,410.5), the number of
items ranged from 3 to 379 (median 19), and the maximum number of categories
ranged from 2 to 11 (median 5). The model split was 120 Rasch and 384 GPCM
units. Stage membership and failure reasons were stored for all 590 attempts.

The IRW is a convenience corpus of data that could be shared. Its unit shares
describe the analyzed corpus and do not estimate frequencies among all
psychological or educational measures.

\begin{table}[H]
\caption{Analysis flow and model composition for the 590 attempted response
matrices.}
\label{tab:osm-flow}
\centering
\small
\begin{tabular}{lrrl}
\toprule
Stage or model family & Units & Studies & Median persons/items/categories \\
\midrule
Candidate units attempted & 590 & --- & --- \\
Worker-level success & 572 & --- & --- \\
EH diagnostic analysis frame & 504 & 273 & 1,410.5 / 19 / 5 \\
Complete normal--EH pair & 496 & --- & --- \\
\addlinespace
Rasch & 120 & 86 & 1,914 / 45 / 2 \\
GPCM & 384 & 196 & 1,375 / 12 / 5 \\
\bottomrule
\end{tabular}
\end{table}

\subsection{Construct Composition}

IRW metadata were collapsed into eight named construct categories and an
unclassified category. \Cref{tab:osm-construct} reports the model composition
and median EH diagnostic in each category. Item-model support was uneven. For
example, the opinion/attitude category contained 4 Rasch and 81 GPCM units,
whereas the cognitive/educational category contained 61 Rasch and 41 GPCM
units. The model-stratified analysis is reported in \cref{app:robust}.

\begin{table}[H]
\caption{Construct composition and median grid-boundary KS distance in the 504-unit
EH diagnostic frame.}
\label{tab:osm-construct}
\centering
\small
\begin{tabular}{lrrrr}
\toprule
Construct & $n$ & Rasch & GPCM & Median KS \\
\midrule
Opinion/attitude & 85 & 4 & 81 & .144 \\
Affective/mental health & 113 & 6 & 107 & .128 \\
Behavioral & 30 & 1 & 29 & .127 \\
Physical health/functioning & 20 & 1 & 19 & .118 \\
Unclassified & 109 & 38 & 71 & .118 \\
Developmental & 8 & 7 & 1 & .101 \\
Other & 9 & 1 & 8 & .096 \\
Cognitive/educational & 102 & 61 & 41 & .074 \\
Personality & 28 & 1 & 27 & .071 \\
\bottomrule
\end{tabular}
\end{table}

%% file: appendices/appendix_C.tex
\section{Model and Calibration Sensitivities}
\label{app:robust}

\subsection{Construct by Item-Model Family}

Four construct groups had at least four units in each model family:
affective/mental health, cognitive/educational, opinion/attitude, and
unclassified. In the pooled additive model, the affective and opinion
contrasts with cognitive/educational were .398 ($p=.007$) and .444
($p=.005$). The corresponding GPCM-only coefficients were .221 ($p=.193$)
and .252 ($p=.150$). The pooled ordering consequently reflects construct and
model composition together. The additive model included a GPCM indicator
($\hat\beta=.036$, $p=.817$), log items, log persons, longitudinal status,
and maximum category count. Standard errors used study-clustered CR2 with
Satterthwaite degrees of freedom over 504 units in 273 studies.

\subsection{Finite-Item Identification Study}

The identification study selected four Rasch and four GPCM units spanning low
and high item counts and low and high EH KS values. Each unit was fitted under
seven quadrature or optimizer-control configurations. Fifty-four of the 56
fits converged. The within-unit KS range varied from .002 to .055.

\begin{figure}[H]
\centering
\includegraphics[width=\textwidth]{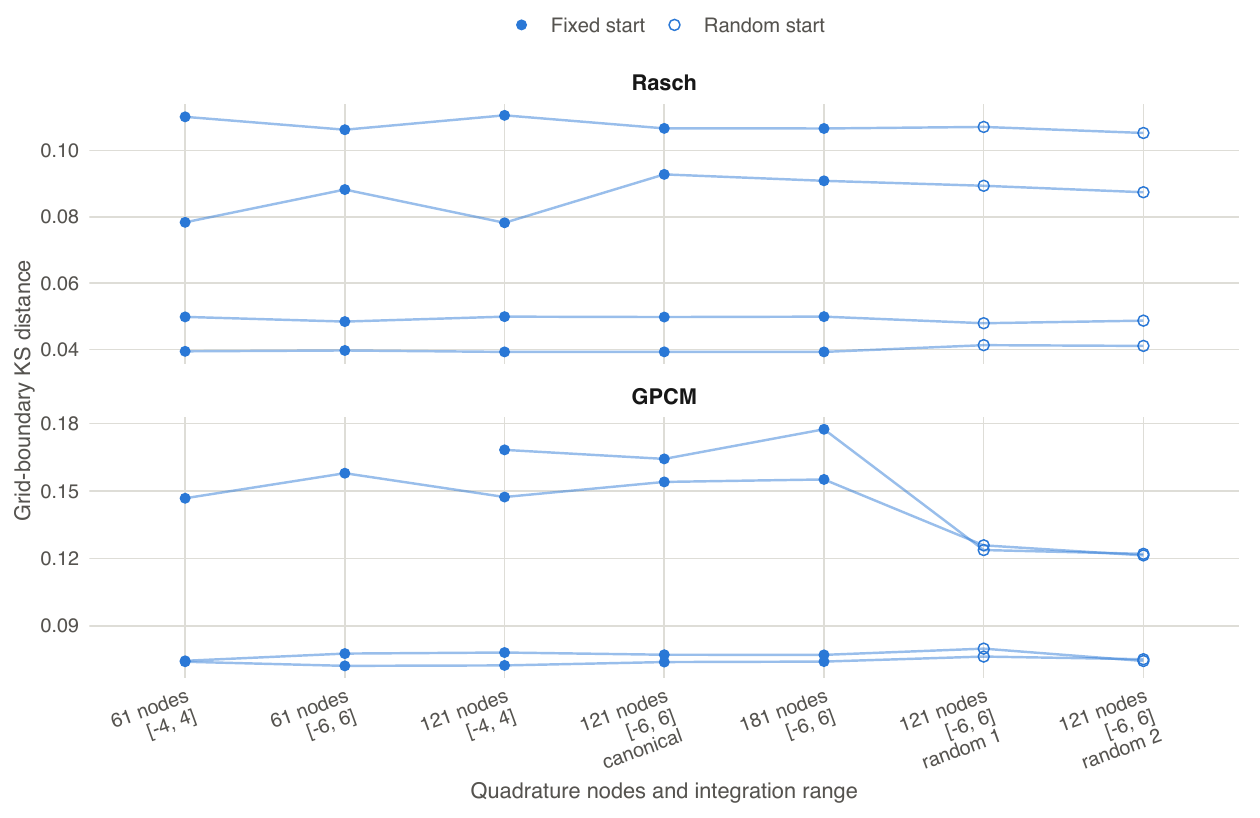}
\caption{Grid-boundary KS distance under seven quadrature and optimizer settings for
eight selected units; 54 of 56 fits converged. Points show fitted values on the
standardized scale, and missing points denote nonconvergence.}
\label{fig:osm-grid}
\end{figure}

The local observed-pattern calculations did not identify a free finite-item
mixing distribution in the selected units. For GPCM, the rank calculation was
conditional on fitted item parameters. For Rasch, the finite response
constraints left degrees of freedom in the grid masses. The numerical ranges
in \cref{fig:osm-grid} summarize stability under nearby estimation settings.

\subsection{Reliability Sensitivity}

\Cref{tab:osm-reliability} reports three absolute between-calibration
differences over the 496 complete normal--EH pairs. Changing only the
integration density while holding item parameters fixed gave a median of
.002. Full recalibration gave a median marginal-reliability difference of
.025 and a median empirical-reliability difference of .005.

\begin{table}[H]
\caption{Absolute reliability differences in 496 complete normal--EH
calibration pairs; eight units in the 504-unit diagnostic frame were missing.}
\label{tab:osm-reliability}
\centering
\small
\begin{tabular}{lrrr}
\toprule
Target & Median & 90th percentile & Share above .05 \\
\midrule
Fixed-item density component & .002 & .019 & .010 \\
Full-refit marginal reliability & .025 & .096 & .272 \\
Full-refit empirical reliability & .005 & .033 & .060 \\
\bottomrule
\end{tabular}
\end{table}

\subsection{Item, Test-Curve, and Score Sensitivity}

The item and score comparisons used the same 496 complete pairs and a common
standardized metric. \Cref{tab:osm-item-score} gives medians and upper
quantiles. Threshold/location RMSD had a long right tail, including a 90th
percentile of 10.371. This tail contains poorly aligned or unstable refits, so
the median and quartiles are the primary summaries.

\begin{table}[H]
\caption{Absolute item, test-curve, and score differences in 496 complete
normal--EH calibration pairs; all quantities use a common standardized metric.}
\label{tab:osm-item-score}
\centering
\small
\begin{tabular}{lrrr}
\toprule
Target & Median & 75th percentile & 90th percentile \\
\midrule
Threshold/location RMSD & .274 & .751 & 10.371 \\
Discrimination RMSD & .161 & .607 & 1.653 \\
Test-curve RMSD per item & .077 & .242 & .497 \\
Person-score mean absolute difference & .093 & .199 & .323 \\
$|z|>1$ crossing fraction & .048 & .113 & .218 \\
\bottomrule
\end{tabular}
\end{table}

The bounded threshold exercise compared the observed normal--EH score
differences with ordinary normal--normal refit differences. Eight units were
used, with two respondent-resampling pairs per unit. At $|z|=1$, the median
crossing fractions were .018 for the observed comparison and .024 for the
resampling comparison. The small replication count makes this a scale
benchmark, not an inferential comparison.

\begin{figure}[H]
\centering
\includegraphics[width=\textwidth]{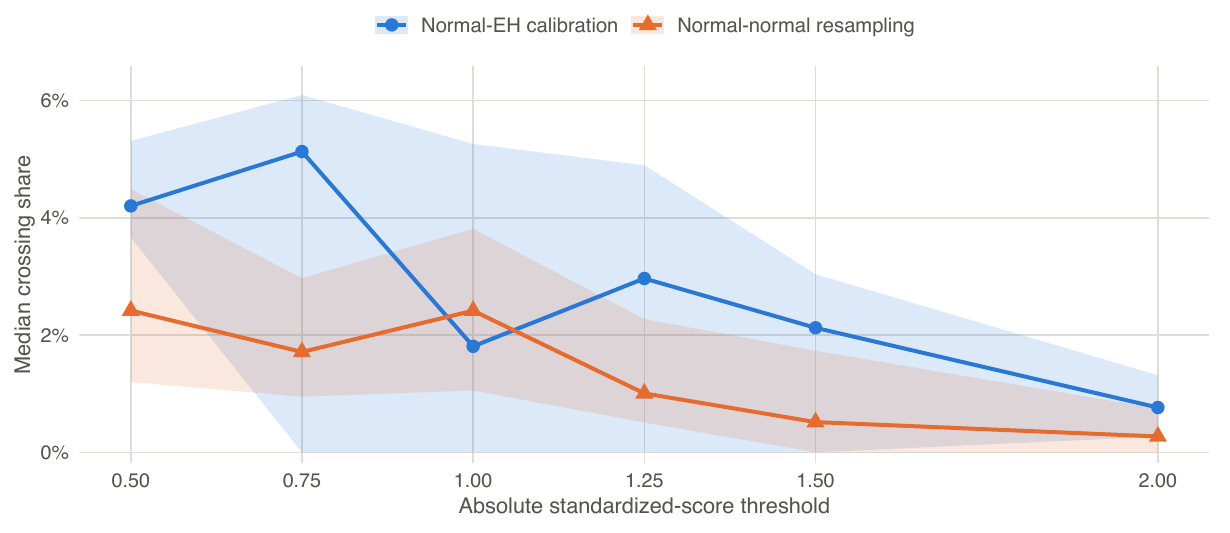}
\caption{Crossing fractions at six standardized boundaries in eight selected
units. The observed series compares normal and EH calibrations; the reference
series uses two normal--normal respondent-resampling pairs per unit.}
\label{fig:osm-threshold}
\end{figure}

\subsection{Item-Response-Function Sensitivity}

Eight dichotomous units were fitted with Rasch, 2PL, and, when available, 3PL
item-response functions under normal and EH mixing distributions. Thirty-two
of 40 attempted fits converged. Among units with both Rasch and 2PL results,
the median absolute change in EH KS was approximately .015. Larger changes
appeared in some 3PL fits, with uneven convergence. These values describe the
joint dependence of the fitted item-response function and mixing distribution.

\begin{figure}[H]
\centering
\includegraphics[width=\textwidth]{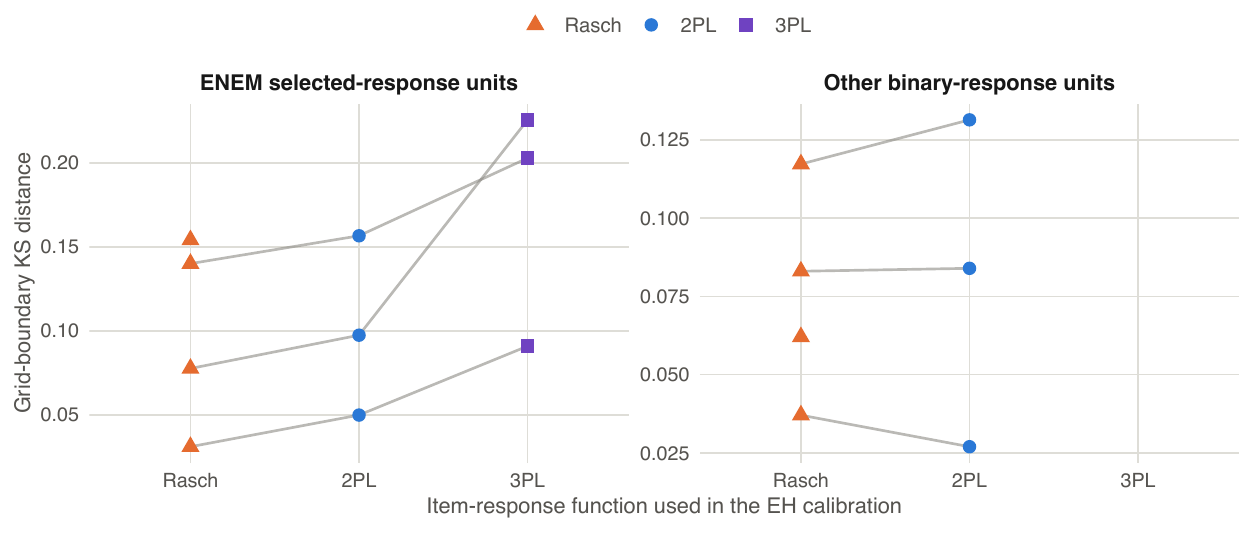}
\caption{Grid-boundary KS distance under Rasch, 2PL, and 3PL item-response functions
for eight dichotomous units; 32 of 40 normal or EH fits converged. Missing
points denote failed or unavailable fits.}
\label{fig:osm-irf}
\end{figure}

%% file: appendices/appendix_D.tex
\section{Density-Estimator Comparison}
\label{app:estimator}

All 504 analysis units were fitted with the extrapolated empirical-histogram
variant implemented in \pkg{mirt} (EHW; \cite{chalmers_mirt_2012}) and with
Davidian curves at candidate orders 2, 4, 6, 8, and 10
\parencite{woods_item_2009}. The resulting estimator
file contained 3,024 unit-by-estimator rows and preserved the revised response
hash for every fit. Candidate order was selected by the Hannan--Quinn criterion
under ceilings of 6 and 10.

EHW converged for 112 of 120 Rasch units and 317 of 384 GPCM units. Davidian
order 2 converged for all units. At the higher orders, 368 to 374 GPCM fits and
116 to 120 Rasch fits converged. Complete-pair comparisons used the
Hannan--Quinn-selected candidate among valid fits.

\begin{table}[H]
\caption{Agreement of density diagnostics across estimators; agreement is the
fraction of complete pairs assigned to the same side of the descriptive
$\KS>.10$ reference.}
\label{tab:osm-estimator}
\centering
\small
\begin{tabular}{lrrrr}
\toprule
Comparison & $n$ & Spearman $\rho$ & Median $|\Delta\mathrm{KS}|$ & Agreement \\
\midrule
EH vs. EHW & 429 & .876 & .009 & .879 \\
GPCM: EH vs. Davidian HQ, cap 6 & 384 & .719 & .045 & .706 \\
GPCM: EH vs. Davidian HQ, cap 10 & 384 & .790 & .042 & .758 \\
GPCM: Davidian cap 6 vs. cap 10 & 384 & .935 & .000 & .938 \\
Rasch: EH vs. Davidian HQ, cap 6 & 120 & .572 & .029 & .750 \\
Rasch: EH vs. Davidian HQ, cap 10 & 120 & .620 & .019 & .808 \\
\bottomrule
\end{tabular}
\end{table}

For GPCM units, the item-model parameterization was held fixed. The EH shares
above .10 were .630 in both Davidian comparisons; the corresponding Davidian
shares were .352 under cap 6 and .409 under cap 10. Increasing the candidate
ceiling improved rank agreement and descriptive-rule agreement. The estimated
diagnostic therefore depends in part on the density family and order-selection
rule.

The available Davidian fit for Rasch units used a common-slope,
fixed-variance parameterization. A two-unit check recovered the common slope,
with KS differences of .001 and .004, but this parameterization differs from
the primary Rasch calibration. The Rasch rows in \cref{tab:osm-estimator} are
reported as joint parameterization sensitivity.

\begin{figure}[H]
\centering
\includegraphics[width=\textwidth]{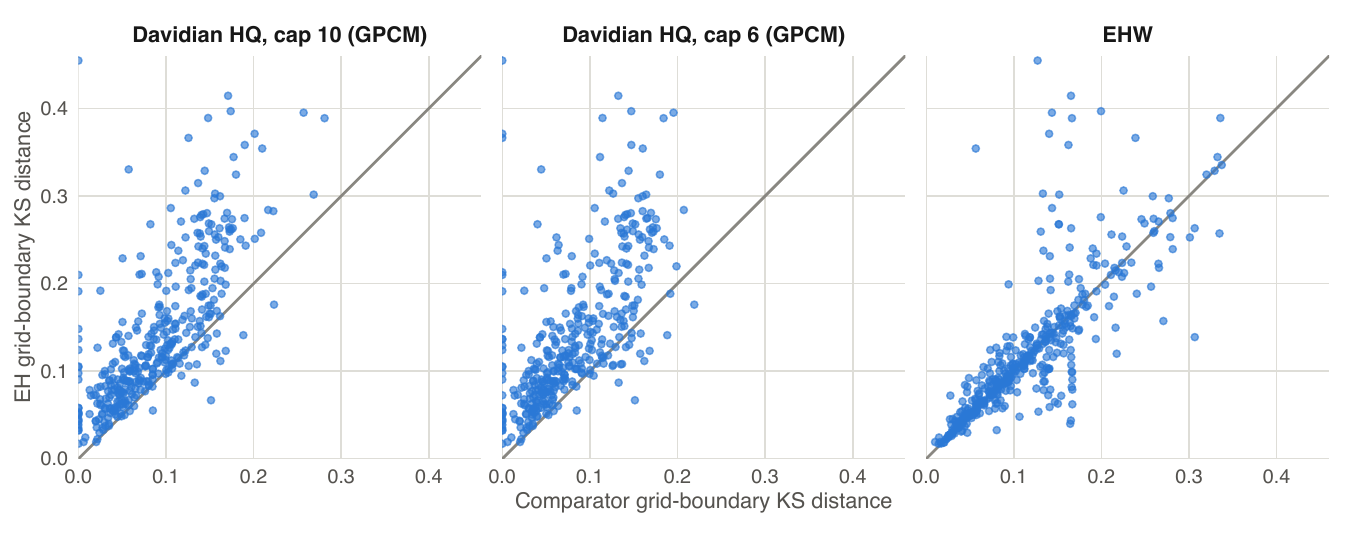}
\caption{Pairwise grid-boundary KS distance comparisons for EH, EHW, and
Hannan--Quinn-selected Davidian estimates. GPCM panels contain 384 complete EH
and Davidian pairs; EHW panels use the available complete pairs.}
\label{fig:osm-estimator}
\end{figure}

Davidian curves can represent multimodal distributions. The known-truth study
in \cref{app:recovery} evaluates the implemented candidate ceilings and
selection rule in the prespecified simulation cells. Its findings concern
those procedures and generating distributions.

%% file: appendices/appendix_E.tex
\section{Known-Truth Density-Estimator Recovery}
\label{app:recovery}

The simulation used a factorial, cell-indexed design following the reporting
structure of \textcite{siepe_simulation_2024}. Four standardized latent shapes
(normal, moderate skew, strong skew, and bimodal) were crossed with 2PL and
GPCM item families and with 10- and 20-item tests. The sample size was 5,000 in
every cell. One hundred Monte Carlo jobs were planned for each of the 16 cells.

Each job fitted EH and Davidian candidates at orders 2, 4, 6, 8, and 10. The
design produced 9,600 estimator-fit rows. Valid cell-by-estimator counts ranged
from 96 to 100. Performance measures were detected-minus-true grid-boundary KS
distance and the KS distance between the estimated and generating distributions.
Monte Carlo standard errors were retained for both measures. The reported
performance targets are density-estimator recovery measures.

\begin{table}[H]
\caption{Mean detected-minus-true grid-boundary KS distance in the 20-item bimodal
cells, based on 100 planned jobs per item family.}
\label{tab:osm-recovery}
\centering
\small
\begin{tabular}{lrr}
\toprule
Estimator & 2PL & GPCM \\
\midrule
EH & .002 & .004 \\
Davidian HQ, candidate cap 6 & $-.099$ & $-.093$ \\
Davidian HQ, candidate cap 10 & $-.042$ & $-.049$ \\
\bottomrule
\end{tabular}
\end{table}

Under normal generation with 20 items, the cap-selected Davidian bias was
approximately zero. EH bias was .012 for 2PL and .009 for GPCM. Under moderate
skew, cap 6 and cap 10 usually selected the same order. Candidate ceiling had
a larger effect under strong skew in some cells and under bimodality in every
cell.

\Cref{tab:osm-recovery} gives the main bimodal comparison at 20 items. EH
detection bias was close to zero. Davidian selection with cap 6 under-detected
the generating KS by .093 to .099. Raising the ceiling to 10 reduced the
magnitude to .042 to .049. The same direction appeared at 10 items, with
larger recovery errors in the GPCM cells.

\begin{figure}[H]
\centering
\includegraphics[width=\textwidth]{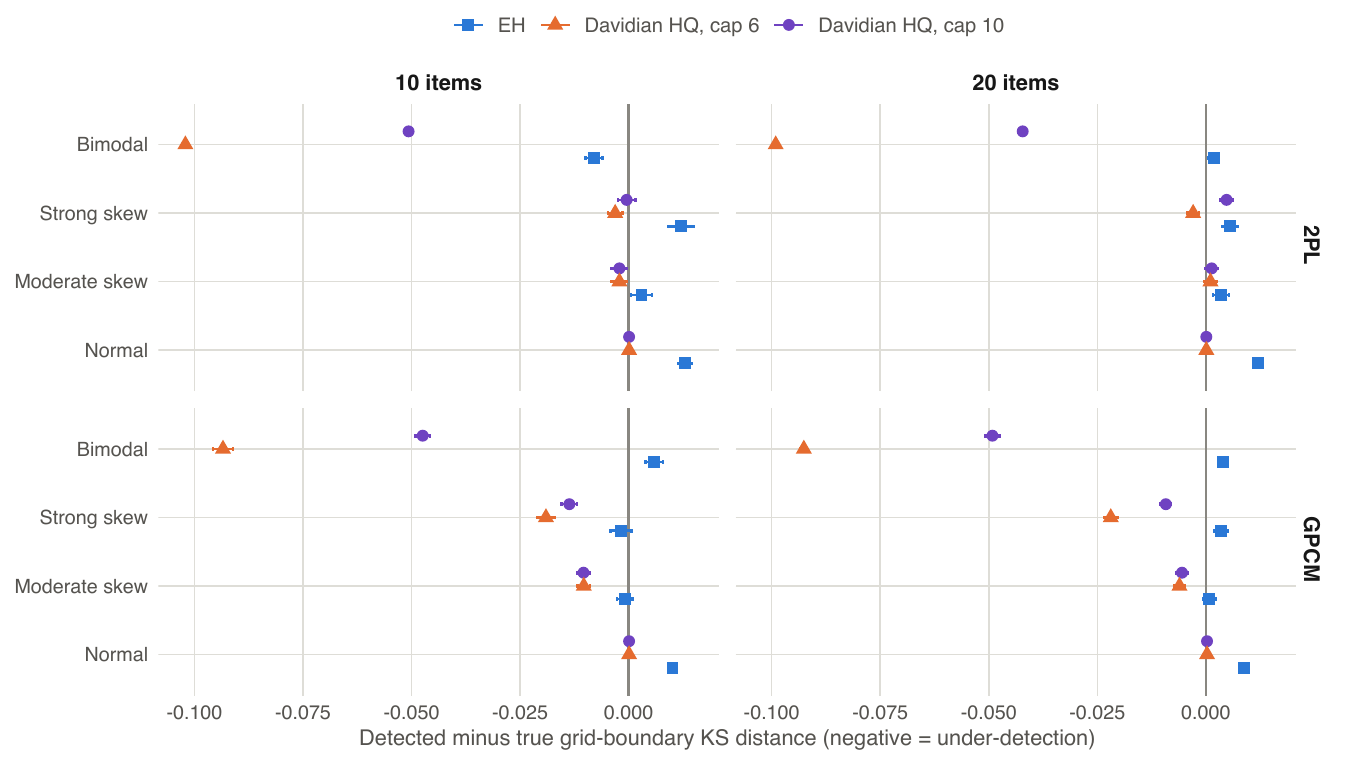}
\caption{Mean detected-minus-true grid-boundary KS distance by generating shape, item
family, test length, and estimator. Each cell planned 100 Monte Carlo jobs;
error bars show Monte Carlo uncertainty from the 96 to 100 valid jobs per
cell-estimator combination.}
\label{fig:osm-recovery}
\end{figure}

The findings apply to the four generating shapes, two item families, two test
lengths, Hannan--Quinn selection, and candidate ceilings used here. They show
that estimator choice can change a descriptive KS magnitude, especially for
the selected bimodal distribution. They do not establish a general ordering
of density estimators for real response matrices.

%% file: appendices/appendix_F.tex
\section{Reproducibility Map}
\label{app:repro}

Each unit record includes its analysis-frame membership,
response hash, person-sampling seed, model and estimator settings, convergence
status, iteration count, and failure reason. The global seed was 20260721.

\begin{table}[H]
\caption{Scripts and primary outputs used for the observed-data corpus,
sensitivity, and density-recovery analyses.}
\label{tab:osm-repro}
\centering
\small
\begin{tabularx}{\textwidth}{@{}l>{\raggedright\arraybackslash}X@{}}
\toprule
Script & Role and output \\
\midrule
76 & Full observed-data normal and EH screen, with one retained record for
each of 590 attempted units \\
77 & Public 134-column dataset at
\code{data-derived/nonbootstrap-dataset.csv} and stage-membership flow \\
100 & Corpus tables, figure-source files, and observed-data figures \\
101 & Unit-matched provenance audit \\
102 & Eight-unit finite-item identification and grid/control sensitivity \\
103 & Construct-by-model summaries and OLS models with study-clustered CR2
uncertainty \\
104--106 & Reliability, threshold, item-response-function, and dimensionality
sensitivity products \\
107 & EHW and five Davidian candidate fits for the 504-unit estimator corpus \\
108 & Known-truth density recovery: 16 cells and 9,600 estimator-fit rows \\
109 & Estimator and recovery tables, source files, and figures \\
\bottomrule
\end{tabularx}
\end{table}

Analyses used R 4.6.0 \parencite{r_core_team_r_2026}, \pkg{mirt} 1.46.1
\parencite{chalmers_mirt_2012}, \pkg{sirt} 4.2.133
\parencite{robitzsch_sirt_2025}, \pkg{diptest} 0.77.2
\parencite{maechler_diptest_2025}, \pkg{clubSandwich} 0.7.0
\parencite{pustejovsky_clubsandwich_2026}, \pkg{data.table} 1.18.4
\parencite{barrett_datatable_2026}, and \pkg{ggplot2} 4.0.3
\parencite{wickham_ggplot2_2016}. \pkg{mirt} fitted the normal, EH, EHW, and
Davidian models. \pkg{sirt} supplied the archived \code{smooth3} comparison
only.

The item-response data are public through the IRW
\parencite{domingue_introduction_2025}. Analysis code, manifests, derived
unit-level summaries, and the mapping from reported results to source runs are
released at \url{https://github.com/joonho112/irw-normality-replication}.